\documentclass[prb,twocolumn,superscriptaddress,showpacs,amsmath,amssymb]{revtex4}

\usepackage{graphicx}
\usepackage{latexsym}
\usepackage{amsmath}
\usepackage{amssymb}
\usepackage{amsfonts}
\usepackage{color}
\usepackage{bm}
\usepackage{verbatim}

\begin{document}

\title{Spin Hanle effect in mesoscopic superconductors}

\author{M. Silaev}
 \affiliation{O.V. Lounasmaa Laboratory, P.O. Box 15100, FI-00076 Aalto University, Finland}

\affiliation{Department of Theoretical Physics, The Royal
Institute of Technology, Stockholm SE-10691, Sweden}

\author{P.~Virtanen}
\affiliation{O.V. Lounasmaa Laboratory, P.O. Box 15100, FI-00076
Aalto University, Finland}

\author{T.T.~Heikkil\"a}

 \affiliation{Department of
Physics and Nanoscience Center, University of Jyv\"askyl\"a, P.O. Box 35 (YFL), FI-40014 University of Jyv\"askyl\"a,
Finland}

\author{F.S.~Bergeret}

\affiliation{ Centro de F\'{i}sica de Materiales (CFM-MPC), Centro
Mixto CSIC-UPV/EHU, Manuel de Lardizabal 5, E-20018 San
Sebasti\'{a}n, Spain}

\affiliation{Donostia International Physics Center (DIPC), Manuel
de Lardizabal 5, E-20018 San Sebasti\'{a}n, Spain}

\affiliation{Institut f\"ur Physik, Carl von Ossietzky
Universit\"at, D-26111 Oldenburg, Germany}

 \begin{abstract}
We present a theoretical study of spin transport in a
superconducting mesoscopic spin valve under the action of a
magnetic field misaligned with respect to the injected spin.
{\color{black} We demonstrate that superconductivity can either
strongly enhance or suppress the coherent spin rotation depending
on the type of spin relaxation mechanism being dominated either by
spin-orbit coupling or spin-flip scattering at impurities. We also
predict a hitherto unknown subgap contribution to the nonlocal
conductance in multiterminal superconducting hybrid structures
which completely eliminates the effect of spin rotation at
sufficiently low temperatures.}
 \end{abstract}

 \maketitle

\section{Introduction}

  Effective control over spin-polarized transport is a cornerstone
  for many spintronics applications \cite{Spintronics}.
   One way of implementing spin manipulation and control is to
  exploit the Hanle effect, i.e., the coherent rotation of a spin in
  an external magnetic field\cite{AwshalomReview}.
Such rotation has been experimentally demonstrated in
semiconducting nanostructures \cite{GaAsSpinRot,SiliconSpinRot},
  graphene \cite{GraphenSpinRot} and normal metals \cite{Johnson1985,Jedema,
  Jedema1,Fukuma2011}.
In the latter case, however, the strong spin-relaxation mechanisms
impose a requirement of much larger magnetic fields to rotate the
spin of rapidly moving electrons within a distance comparable to
the spin coherence length \cite{NazarovNonCollinear}.

{ One possible alternative to tune the spin rotation in metals is
to use the intriguing spin transport properties of superconductors
\cite{HanleSuper,Hubler,Aprili} which have stimulated recently a
broad interest and rapid progress in the emergent field of
superconducting
spintronics\cite{ScientificReports,EschrigPhysToday,
BergeretNatureComm2014}. In contrast to the normal metals
 the spin in superconductors is transported by
 the Bogoliubov quasiparticles}, which move at the
 group velocity $v_g\sim v_F \sqrt{(\varepsilon/\Delta)^2-1}$ that tends
 to zero near the gap edge $\varepsilon=\Delta$. Hence while travelling a
 fixed distance $L$ they are exposed  to the spin-rotating field
 for a longer time, which results in an enhanced spin precession.
 This also implies an increase of
 the spin relaxation due to the exchange interaction with magnetic
 impurities\cite{morten04,morten05,Poli}.
  However, besides generating the spin rotation and relaxation,
the magnetic field induces a Zeeman splitting of quasiparticle
states which can lead to the separation of spin and charge degrees
of freedom \cite{SpinChargeSeparation} and a drastic suppression
of spin relaxation \cite{Aprili,Hubler,SpinInjectionNb}. Thus, in
principle the magnetic field can either enhance the spin
relaxation due to the Hanle mechanism or suppress it by the
polarization of the quasiparticles.  Although some works have been
devoted to the theory of spin relaxation in superconductors
\cite{morten04,morten05,Poli,SpinChargeSeparation}, none of them
have addressed the problem of noncollinear spin-splitting fields,
essential to understand   the Hanle effect in superconductors.

 In this paper we address this problem and  present a full study of  spin transport in  a typical  nonlocal measurement setup.
   We extend the existing spin transport theory in diffusive
 superconductors \cite{morten04,morten05} by taking into account  noncollinear configurations
of spin injector and detector electrodes and an external magnetic
field in  an arbitrary direction. We show that the nonlocal
magnetoresistance depends crucially  on the spin relaxation
mechanism in the superconductor.
 If the latter is mainly due to an extrinsic
spin-orbit coupling, the nonlocal spin signal in the
superconducting state is suppressed by smaller fields as compared
to the normal case, whereas the period of characteristic
oscillations of the Hanle curve becomes smaller. In contrast, if
the main source of spin relaxation is due to magnetic impurities
(spin-flip scattering) the decay of the nonlocal spin signal with
the applied field is larger in the superconducting state but its
oscillation become less pronounced for $T<T_c$. Our theory also
predicts that  the injection of spins noncollinear with the
external
 field  can generate in the superconductor a subgap pure spin imbalance that
 provides an additional, and  hitherto unknown contribution to the subgap
 nonlocal conductance  in multiterminal superconducting hybrid structures\cite{HekkinkCAR,BeckmanCAR,ZaikinCAR}.

The structure of this paper is as follows. In Sec. \ref{Sec:Model}
 we describe a model of the non-local resistance measurements and the general Usadel theory and
 in spin-polarized diffusive superconductors.
In Sec. \ref{Sec:SpinCoherence} we derive the equations describing
 spin injection, rotation and decay of spin coherence in diffusive superconductors
 under the action of a transverse magnetic field. In Sec. \ref{Sec:HanleEffect}
 we discuss in detail the Hanle effect and give the conclusions in
 Sec. \ref{Sec:Conclusion}.

\section{Model} \label{Sec:Model}

\subsection{Nonlocal spin valve}

 \begin{figure}[!htb]
 \centerline{\includegraphics[width=1.0\linewidth]{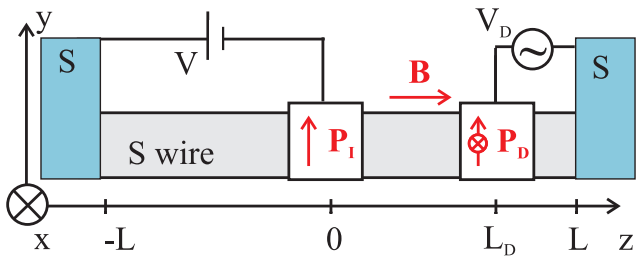}}
 \caption{\label{Fig:Sketch} (Color online) Schematic view of a  nonlocal conductance measurement setup
 with noncollinear polarization of ferromagnetic contacts and magnetic field.
 We choose the injector polarization  ${\bm P}_I\parallel{\bm y}$,
 the magnetic field  ${\bm B}\parallel {\bm z}$, while
 the detector polarization ${\bm P}_D$ has an arbitrary direction. }
 \end{figure}

We consider the nonlocal spin valve shown in Fig.
\ref{Fig:Sketch}. A spin-polarized current is injected in the
superconducting wire from a ferromagnetic electrode with
magnetization ${\bm P}_I$. The detector is also  a ferromagnet
with a polarization ${\bm P}_D$ located at a distance $L_D$ from
the injector. Both the injector and the detector are coupled to
the wire via tunnel contacts. A magnetic field ${\bm B}$ is
applied in $z$ direction. The current $I_D$ at the detector is
given by (below, we use $\hbar=e=k_B=1$)
 \begin{equation}\label{Eq:ZeroCurrentYGen}
 R_DI_D= \mu_0   + \bm \mu\cdot \bm {P}_D,
 \end{equation}
 where $R_D$ is the detector interface resistance in the normal state, $\mu_0$ is the effective charge imbalance
 of quasiparticles in the superconductor, and the last term describes the spin dependent contribution to the current which
 is proportional to the local spin accumulation ${\bm \mu}$. Expression (\ref{Eq:ZeroCurrentYGen}) follows directly from the boundary conditions at the
 spin-polarized ferromagnet/superconductor interface as we show in Sec. \ref{Sec:BC}.

 In a nonlocal measurement scheme,  the spin accumulation  is tested by measuring the voltage at the detector where
 no charge current flows \cite{Hubler,Poli}, {\it i.e.}, one sets in Eq.~\eqref{Eq:ZeroCurrentYGen} $I_D=0$.
 The spin-dependent voltage $V_S$ is defined as the difference of voltages measured in the parallel and anti-parallel configurations
 between the injector and the detector.
 The nonlocal spin signal is determined by the ratio
 $R_S=V_S/I_{inj}$, where $I_{inj}=V\chi/R_I$ is the injected current,
 $R_I$ is the injector interface resistance, and $\chi=\int_{0}^{\infty} d\varepsilon N_+\frac{\partial n_0}{\partial\varepsilon}$
 is the "Yosida function"\cite{Poli}.  Here $N_+$
 is the density of states (DOS) in the superconductor near the ferromagnetic electrode and
 $n_0(\epsilon)$ is the Fermi distribution function. As usual, we consider only
 the linear response limit $|V| \ll T$.
 The divergence of the spin signal in the low-temperature limit \cite{SpinSignalDivergence} is cut off
 by proximity-induced subgap contributions to $N_+$. The corresponding nonlocal resistance
is
 \begin{equation}\label{Eq:SpinSignal}
 R_{S} = 2R_I(V\chi^2)^{-1} ({\bm \mu\cdot \bm{P_D}})\;,
 \end{equation}
as obtained from Eq. (\ref{Eq:ZeroCurrentYGen}). Below we
calculate in detail the spin accumulation ${\bm \mu}$ in a
superconductor with Zeeman splitting, and discuss the resulting
behavior of $R_S$.

 \subsection{ Non-equilibrium Green's functions and the Usadel equation}

 The  spin accumulation ${\bm \mu}$ in Eqs. (\ref{Eq:ZeroCurrentYGen},\ref{Eq:SpinSignal})
  can be written in terms of the Keldysh quasiclassical Green's function (GF) as
   \begin{equation}
 {\bm \mu} =  \int_{0}^\infty \bm m(\varepsilon) d\varepsilon \; ,
 \end{equation}
  where ${\bm m}(\varepsilon) = {\rm Tr}\; (\tau_3 \bm \sigma g^K ) /8$, $\tau_3$ is the third Pauli matrix in Nambu space,
 $g^K$ is the (2$\times$2 matrix) Keldysh component of the quasiclassical GF matrix
 \begin{equation}
 \check{g} = \left(%
 \begin{array}{cc}
  g^R &  g^K \\
  0 &  g^A \\
 \end{array}\label{eq:GF0}
 \right)\; ,
 \end{equation}
 and $g^{R(A)}$ is the retarded (advanced) GF. In a diffusive superconducting wire the matrix $\check g$
 obeys the Usadel equation \cite{Bergeret2001}
  \begin{equation}\label{Eq:Usadel1}
 \frac{D}{2}\nabla\cdot(\check{g}\nabla\check{g})+ [\check\Lambda - \check\Sigma_{so} - \check\Sigma_{sf}, \check{g}] =0.
 \end{equation}
   Here $D$ is the diffusion constant, $\check\Lambda = i\varepsilon \tau_3-i({\bm{ h\cdot S})}\tau_3 - \check{\Delta}$, $\varepsilon$ is the energy,
   $ \check{\Delta}=\Delta\tau_1$ the spatially homogeneous order parameter in the wire,  ${\bm h } = \mu_B {\bm B}$ the Zeeman field,
   $\mu_B$ the Bohr magneton, and ${\bm S}= (\sigma_1,\sigma_2,\sigma_3)$ the vector of Pauli matrices in spin space. The last two terms in
   Eq.~\eqref{Eq:Usadel1},  $\check\Sigma_{so} = \tau_{so}^{-1} ({\bm {S}}\cdot\check{g} {\bm {S}})$ and
   $\check\Sigma_{sf} = \tau_{sf}^{-1} ({\bm {S}}\cdot\tau_3\check{g} \tau_3 {\bm {S}})$,
 describe spin relaxation due to
 spin-orbit scattering and exchange interaction with magnetic
 impurities, characterized by relaxation times $\tau_{so}$ and
 $\tau_{sf}$,
 respectively. The commutator is defined as $[A,B]=(AB-BA)/2$.  Equation \eqref{Eq:Usadel1} is complemented by the normalization condition $\check g^2=1$ that allows writing the Keldysh component as
 $ g^K= g^R  \hat f - \hat f  g^A$, where  $\hat f$ is the distribution function  with a general spin structure
  \begin{equation}\label{Eq:DistrFun}
   \hat f= f_L +f_T\tau_3  + ({\bm \sigma \cdot \bm {f_T}})+  ({ \bm \sigma \cdot \bm {f_L}}) \tau_3.
  \end{equation}
  {\color{black} Here the L-labeled functions denote the (spin) energy degrees of freedom and are always
  antisymmetric with respect to the Fermi level $\varepsilon=0$ in
  the superconductor. The T-labeled functions are symmetric and describe the charge/spin imbalance. This is a generalization of the Schmid-Sch\"on
  theory to the case of non-collinear spin transport\cite{morten04,morten05,SchmidSchon}. }

 \begin{figure}[!htb]
 \centerline{\includegraphics[width=1.0\linewidth]{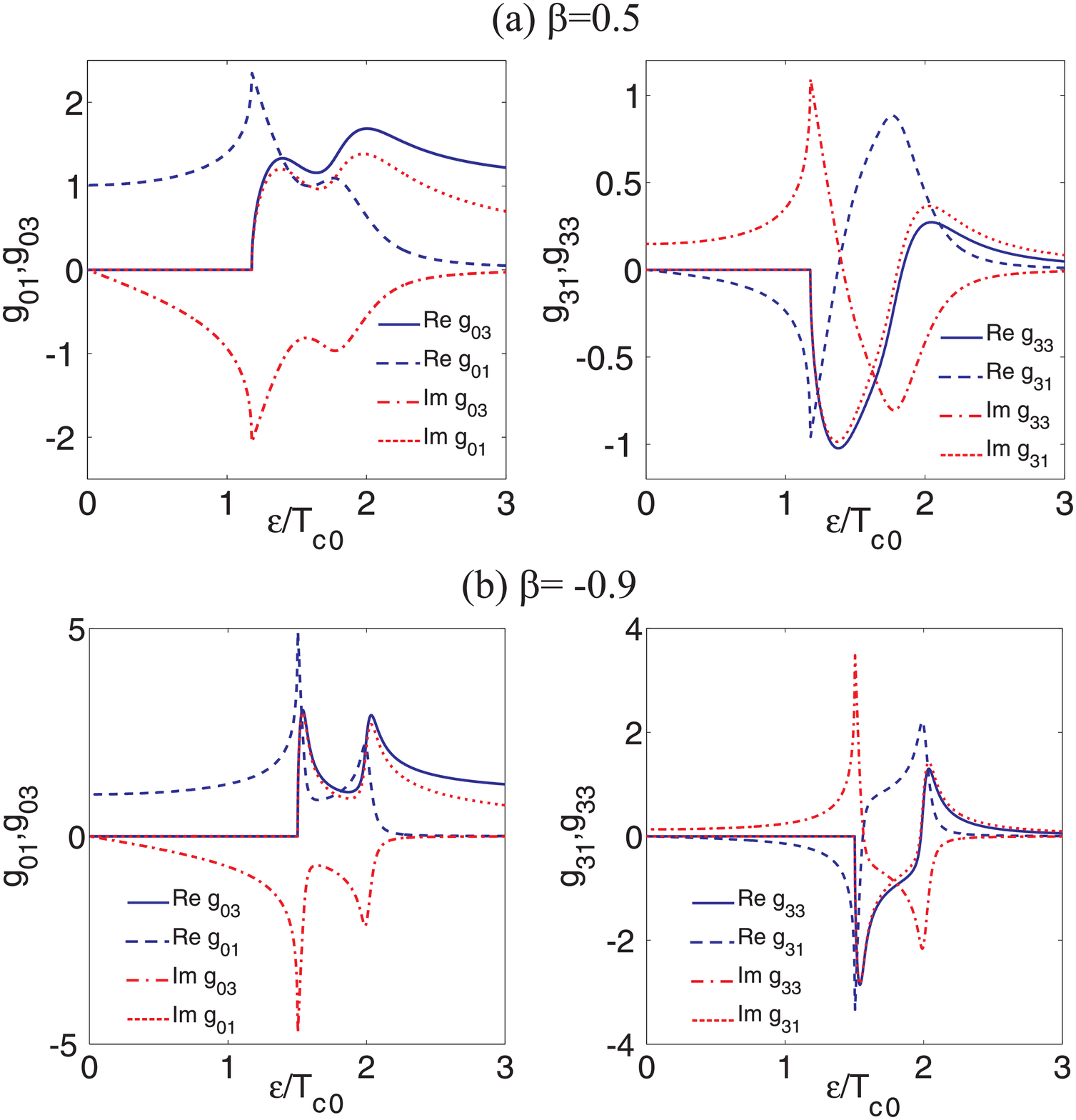}}
 \caption{\label{Fig:GF} (Color online) Components of the spectral
 GF $g^R$ for the different relative strengths of spin-flip and spin-orbital scattering (a) $\beta
 =0.5$ (dominating spin-flip scattering) and (b) $\beta=-0.9$ (dominating
 spin-orbital scattering). Left and right panels show spin singlet and spin triplet GF components correspondingly.
  The component $N_+ = {\rm Re} g_{03}$ is the total DOS in spin
  subbands. The parameters are $T=0.25 T_{c0}$, $\tau_{sn}=5/T_{c0}$, and $h=10/\tau_\Sigma$. }
 \end{figure}

 We assume that the transparencies of the detector and injector interfaces are small, so that up to
  leading order the spectral (retarded and advanced) GFs obtain their bulk values in the presence
  of a  Zeeman splitting field.  In the present case  ${\bm h} = h{\bm z}$,  they read
 \begin{equation} \label{Eq:GR}
 g^R = \tau_1(g_{01}+ g_{31} \sigma_3) +\tau_3( g_{03}  + g_{33} \sigma_3)\; ,
 \end{equation}
 and $ g^A=-\tau_3 g^{R\dag}\tau_3$.
 While the terms diagonal in Nambu space ($\tau_3$) correspond to the normal GFs,
   $g_{01},g_{31}$ are the singlet and zero-spin triplet
 anomalous components which describe the  superconducting condensate \cite{Bergeret2001}.
 The coefficients in Eq.~\eqref{Eq:GR} are determined by solving the nonlinear
 equation
 \begin{equation}
  [\Lambda^R-\Sigma^R_{sf}-\Sigma^R_{so}, g^R]=0\; ,
  \end{equation}
  where the spin-dependent scattering self-energies
 reduce the (self-consistent) spectral gap to $\Delta_g<\Delta-h$
 and smear the gap edge singularities  \cite{AbrikosovGorkov}, as shown in Fig.~\ref{Fig:GF}.
 Apart from a temperature interval in the vicinity of $T_c$, the applied fields are much smaller than the
 paramagnetic Chandrasekhar-Clogston limit\cite{SaintJames} $|h|\ll \Delta/\sqrt{2}$.

 \subsection{Currents and boundary conditions}\label{Sec:BC}

In terms of the GF, Eq. (\ref{eq:GF0}),
the bulk currents are given by
 \begin{equation}\label{Eq:CurrentBulk}
 {\bm j}_{ki}=\frac{\sigma_N}{8} \int_{0}^{\infty}
 d\varepsilon\;
 {\rm Tr}\; \tau_k \sigma_i (\check{g}\nabla \check{g})^K,
 \end{equation}
 where $\sigma_N$ is the normal state conductivity, $k=0,3$, $i=0,1,2,3$ and the sets of
 indices $(k=3; i=0)$, $(k=0; i=0)$, $(k=0; i=1,2,3$) and $(k=3; i=1,2,3)$  correspond to the charge, energy, spin, spin energy
 currents respectively.

The GF has to be determined from
the Usadel equation (\ref{Eq:Usadel1}) which is completed by boundary conditions
(BC) at the spin-polarized injector interface $z=0$. We use here the BC
of Ref.~\onlinecite{bergeret12} that generalizes the Kupriyanov-Lukichev
\cite{KL} one to the case of spin-dependent barrier transmission.
These BC are obtained by matching the bulk currents in the superconductor with the tunneling currents
through the interface.

  The currents through a ferromagnetic barrier
 characterized by an effective polarization ${\bm P}=P{\bm z}$
are given by\cite{bergeret12}
 \begin{equation}\label{Eq:CurrentTunnellSN}
 {\bm j}^t_{ki}=\frac{1}{8R_{\square}} \int_{0}^{\infty} d\varepsilon
 \;{\rm Tr}\; \tau_k \sigma_i [\hat\Gamma\check{g}_N\hat\Gamma^+, \check{g} ]^K.
 \end{equation}
Here $R_{\square}$ is the barrier resistance per unit area,
$\check{g}_N$ is the  matrix GF of the
 normal electrode. The components of $\check{g}_N$ are given by
 $g^{R(A)}_N= \pm \sigma_0\tau_3$ and $g^{K}_N=2 (n_-+\tau_3
 n_+)$, where $n_{\pm}=[n_0(\varepsilon+V)\pm
 n_0(\varepsilon-V)]/2$ is the voltage-biased distribution
 function in the normal metal electrode, $n_0(\varepsilon)=
 \tanh(\varepsilon/2T)$.
The spin-filtering tunneling matrix $\hat \Gamma= t \tau_3+
u\sigma_3$ is defined through the normalized
 transparencies $t^2+u^2=1$ which satisfy the condition $ut=P/2$.

Throughout this work we assume that the applied field is in $z$
direction, while the magnetization of the injector and detector
may point in another direction. Noncollinear polarization of the
barrier, e.g. rotated by the angle $\alpha$ in the $yz$ plane,
${\bm P}=P(\cos\alpha\; {\bm z} + \sin\alpha\; {\bm y})$,
corresponds to spin rotation of the GF in the superconductor
around the $x$ axis
 \begin{eqnarray}
  \check{g}=\hat R^+ \check{g}_{new} \hat R
 \label{Eq:ExchFieldRotation}
\\
 \hat R= e^{i\alpha\sigma_1/2},
 \end{eqnarray}
 which modifies the transparency matrix as $\hat{\tilde{\Gamma}}=\hat R\hat\Gamma $.
Equation~(\ref{Eq:ZeroCurrentYGen})
for the current measured by the detector electrode follows directly from
Eqs.~(\ref{Eq:CurrentTunnellSN},\ref{Eq:ExchFieldRotation}) by choosing ${\bm P} = {\bm
 P}_D$ and $R_D=R_\square/A_D$, where $A_D$ is the effective
area of the detector.

 The BC at the injector electrode are obtained from the
 conservation of the spectral current density $ {\bf j}_{ki}=  {\bf j}^t_{ki}$
 and have the form
   \begin{equation}\label{Eq:BC1}
  \check{g}\nabla \check{g} =
  \kappa_I [\tilde{\check{g}}_N, \check{g}],
  \end{equation}
  where $\kappa_I=1/(R_{I\square}\sigma_N)$ is the injector
transparency and $\tilde{\check{g}}_N=\hat{\tilde{\Gamma}}
\check{g}_N  \hat{\tilde{\Gamma}}$. Let us consider the
 matrix BC Eq. (\ref{Eq:BC1}) in components. The R,A elements yield
   \begin{eqnarray}\label{Eq:BCSpectralR}
   g^R\nabla g^R = \kappa_I [\tilde{g}^R_N, g^R]  \\ \label{Eq:BCSpectralA}
   g^A\nabla g^A = \kappa_I [\tilde{g}^A_N, g^A]
  \end{eqnarray}
 and the Keldysh component can be written as follows
 \begin{eqnarray}
 \nabla f - g^R\nabla f g^A + g^R\nabla g^R f -f g^A\nabla g^A =
 \\\nonumber
  \kappa_I \left( \tilde{g}^R_N g^K + \tilde{g}^K_N g^A - g^R\tilde{g}^K_N - g^K\tilde{g}^A_N
  \right)/2.
 \end{eqnarray}
 With the help of Eqs. (\ref{Eq:BCSpectralR},\ref{Eq:BCSpectralA}) the above expression can be
 simplified:
   \begin{eqnarray}\label{Eq:BCdf} \nonumber
  & \nabla f - g^R\nabla f g^A = \\
  &\kappa_I ( \tilde{g}^R_N  \tilde{f} g^A   +  g^R \tilde{f} \tilde{g}^A_N
   - g^R\tilde{g}^R_N \tilde{f} - \tilde{f}\tilde{g}^A_N g^A )/2,
  \end{eqnarray}
  where $ \tilde{f} = f_N - f$ is the difference between  distribution functions in the superconducting wire
  (\ref{Eq:DistrFun}) and in the normal metal electrode $f_N = n_+ + n_-\tau_3$.

  In the next sections we discuss the  relaxation and precession of the spin in
  a diffusive superconductor, by solving the boundary problem
  described by equations  (\ref{Eq:Usadel1},\ref{Eq:BCSpectralR},\ref{Eq:BCSpectralA},\ref{Eq:BCdf}).

  \section{Spin precession and relaxation in a diffusive superconductor} \label{Sec:SpinCoherence}

  Although the coherent precession  of spin has been studied extensively in a number of semiconducting and normal metal systems
  \cite{GaAsSpinRot,SiliconSpinRot,GraphenSpinRot,Johnson1985,Jedema, Jedema1,Fukuma2011},
  a theory for superconductors is still lacking.
  In this section we  derive from the general expressions presented in the previous section,
  a compact set of kinetic equations and boundary conditions which
  describe the injection of a transverse spin polarization, its precession and relaxation.
   In the next section we solve these equations and address the spin precession measurable
   in non-local spin valves (see Fig. \ref{Fig:Sketch}), known as the  Hanle effect.

    For simplicity,  we fix the directions of  the injecting
  electrode polarization to ${\bm P_I}=P_I{\bm y}$ and the Zeeman field ${\bm h}=h{\bm z}$,
  and allow the polarization of the detector to have an arbitrary direction.
  In this case $\mu_z=0$, and to evaluate the non-local electric signal (\ref{Eq:ZeroCurrentYGen})
  it is sufficient to consider only a transversal component of the spectral spin polarization
   ${\bf m}_\perp=(m_x,m_y,0)$, which is given by
  \begin{equation}
  \label{Eq:Magnetization}
  {\bm m}_\perp =  N_+ {\bm {f}}_T  + h^{-1}{\rm Im} g_{33}( {\bm{f}_T\times \bm h }),
  \end{equation}
   where ${\bm {f}}_T=(f_{T1}, f_{T2},0)$.
   The first term in Eq.~(\ref{Eq:Magnetization}) describes the quasiparticle contribution.
   It is proportional to the total DOS $N_+={\rm Re} g_{03}$
   modified by the Zeeman splitting and the
   spin dependent scattering mechanisms (see Fig.~\ref{Fig:GF}, left panels). This contribution is only finite for
   energies above the spectral gap $\varepsilon>\Delta_g$.
   The second term in Eq.~\eqref{Eq:Magnetization}, being proportional to  ${\rm Im} g_{33}$,
    is nonzero only if the superconducting spectrum is spin-polarized.
   In contrast to the usual quasiparticle contribution,
   this term is not suppressed at low temperatures $T\ll T_c$ since ${\rm Im} g_{33}$ is nonzero at
   subgap energies (dash-dotted curve in
   Fig. \ref{Fig:GF}b).  As we demonstrate below, this term
  leads to a finite subgap nonlocal
  conductance in the lowest order in transparency. This contribution only involves
  spin degrees of freedom, in contrast to previous works on the subgap charge transport
  \cite{HekkinkCAR,BeckmanCAR,ZaikinCAR}.

   Let us derive the equations for the transverse distribution function
   ${\bm {f}}_T$ from the general Eq.(\ref{Eq:Usadel1}).
   These components are decoupled from the others and satisfy
   the kinetic equations
  \begin{eqnarray} \label{Eq:KinfT12-1}
  {\mathcal{D}}_{T1}\nabla^2 f_{T1}+{\mathcal{D}}_{T2}\nabla^2 f_{T2} &=&  X_1 f_{T1} + X_2 f_{T2}
  \\ \label{Eq:KinfT12-2}
  {\mathcal{D}}_{T1}\nabla^2 f_{T2}-{\mathcal{D}}_{T2}\nabla^2 f_{T1} &=&  X_1 f_{T2} -X_2
  f_{T1},
  \end{eqnarray}
    where the renormalized diffusion coefficients
    are given by
    \begin{eqnarray} \label{Eq:Diffusion-1}
   {\mathcal{D}}_{T1} &=& 1+|g_{03}|^2-|g_{01}|^2+|g_{31}|^2-|g_{33}|^2
   \\ \label{Eq:Diffusion-2}
   {\mathcal{D}}_{T2} &=& 2 \; {\rm Im}\;\left(  g_{33}g^*_{03} - g_{31}g^*_{01}  \right).
   \end{eqnarray}
  In Eqs.~(\ref{Eq:KinfT12-1},\ref{Eq:KinfT12-2}) we have defined  $X_1=(S_{T1} - H_1)/D$ and $X_2=(S_{T2} + H_2)/D$, where
  \begin{eqnarray}\label{Eq:Hanle}
      H_1 &=& 4 h{\rm Im} g_{33},\;\;\; H_2= 4 h N_+ , \\ \label{Eq:S1}
    S_{T1} &=& 2\tau_{sn}^{-1}[ ({\rm Re} g_{03})^2 +  \beta({\rm Im} g_{01})^2],
    \\\label{Eq:S2}
    S_{T2} &=& 2\tau_{sn}^{-1} ( {\rm Im} g_{33}{\rm Re} g_{03} - \beta{\rm Im} g_{01}{\rm Re} g_{31}  )
  \end{eqnarray}
   Here the 'Hanle' terms, (\ref{Eq:Hanle}), describe coherent spin
 rotation and relaxation due to the randomization of spin
 precession phase under the action of the external
 magnetic field. Another source of spin relaxation are the elastic spin-flip and spin-orbital scattering described by
the terms (\ref{Eq:S1},\ref{Eq:S2}). These terms are determined by two parameters which are
   the normal state spin relaxation time
 $\tau_{sn}=\tau_{sf}\tau_{so}/8(\tau_{sf}+\tau_{so})$
and the relative strength of spin-orbital and spin-flip
  scattering $\beta= (\tau_{so}-\tau_{sf})/(\tau_{so}+\tau_{sf})$. For example, in {\rm Al} wires used in
  the spin-transport experiments, the typical spin relaxation time is $\tau_{sn} \approx 100$ ps $\approx
  5/T_{c0}$,
  where $T_{c0} \approx 1.6$ K is the bare critical temperature of
  the superconductor in the absence of the exchange field\cite{Poli}.
  In Al, $\beta=0.5$ indicating the dominating spin-flip relaxation mechanism\cite{Poli}
  while for Nb, one expects spin-orbit as the main source of scattering\cite{SpinInjectionNb}.

\begin{figure}[!htb]
 \centerline{\includegraphics[width=1.0\linewidth]{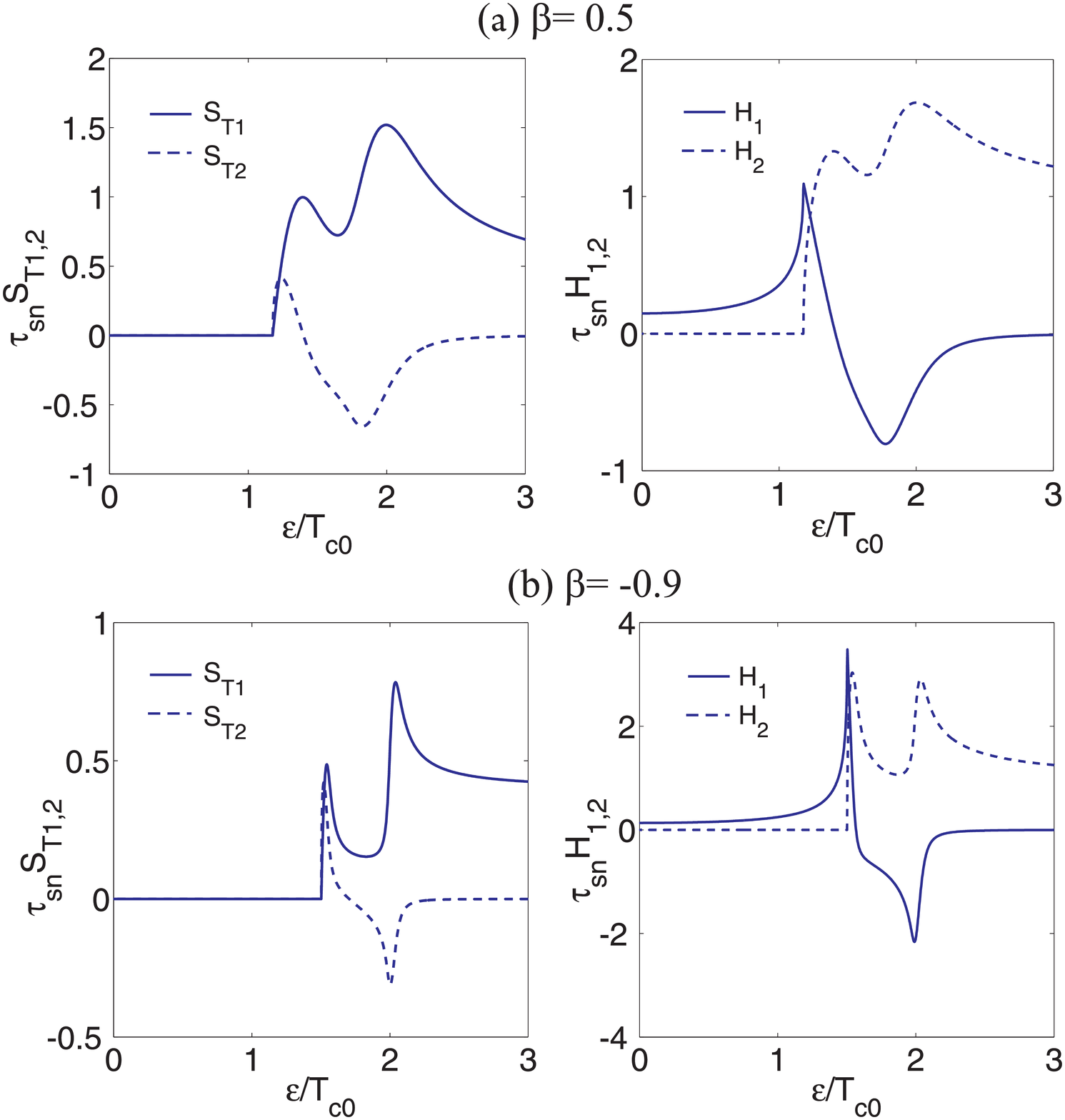}}
 \caption{\label{Fig:SH} (Color online)
 Energy dependencies of the spin relaxation terms $S_{1,2}$ (left panels) and Hanle terms $H_{1,2}$
 (right panels) for (a) $\beta=0.5$ and (b) $\beta=-0.9$. }
 \end{figure}

\begin{figure}[!htb]
 \centerline{\includegraphics[width=1.0\linewidth]{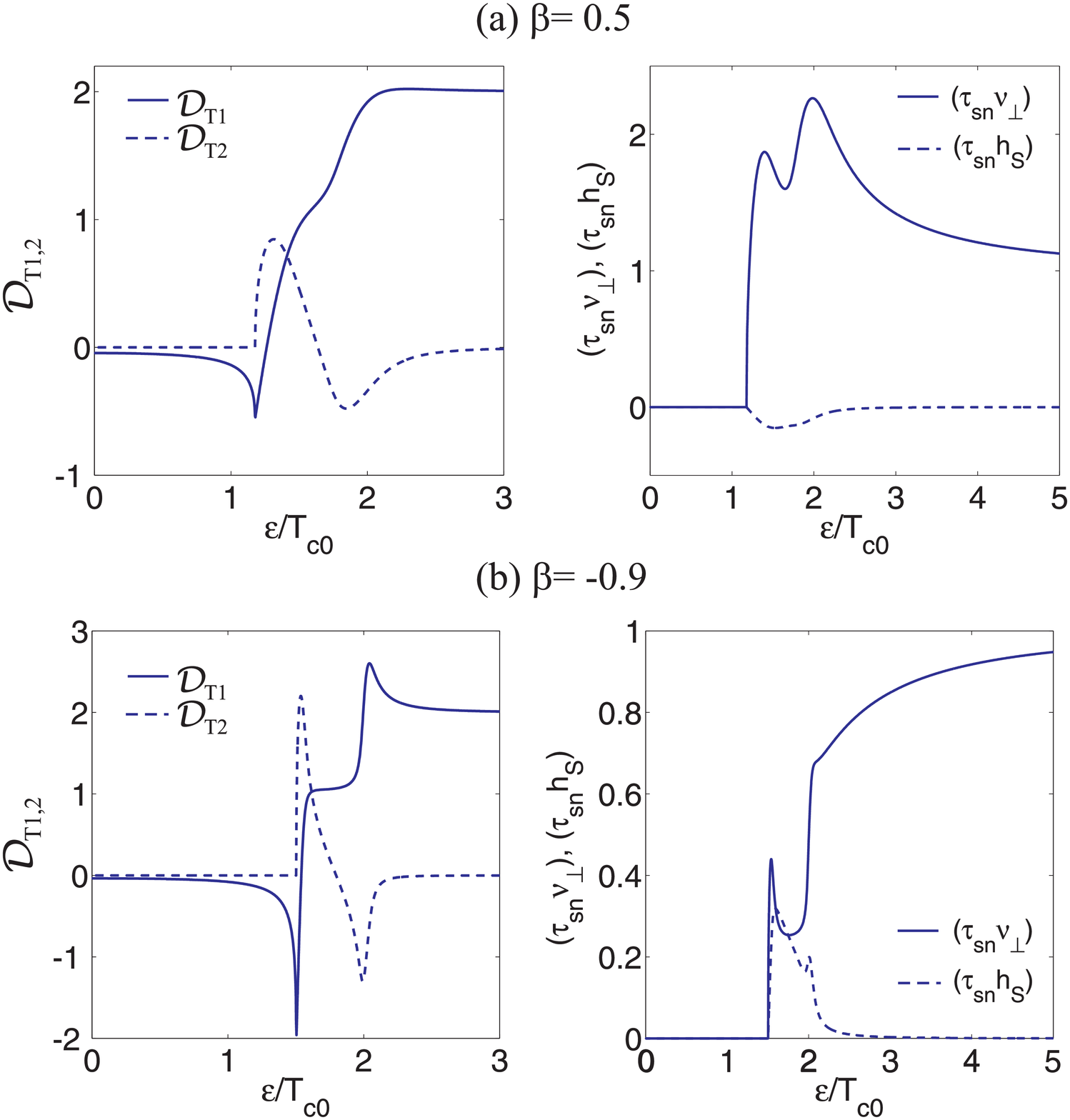}}
 \caption{\label{Fig:DOm} (Color online)
 Energy dependencies of the diffusion coefficients ${\cal D}_{T1,2}$ (left panels)
 and the characteristic frequencies of the spin diffusion Eq.(\ref{LL}) which are the spin coherence decay rate
 $\nu_\perp=1/\tau_\perp$ and the precession frequency shift
 $h_s$. The plots are shown for (a) $\beta=0.5$ and (b) $\beta=-0.9$. }
 \end{figure}

  It is important for the discussions below to understand the energy dependencies of the spin relaxation,
   and Hanle terms,
  which are shown in Fig.~(\ref{Fig:SH}) for spin-flip $\beta=0.5$ (a)  and
  spin-orbital $\beta=-0.9$ (b) dominated scattering.
  Correspondingly we show in the left panels of  Fig.(\ref{Fig:DOm})
  the dependencies of  the diffusion coefficients ${\cal D}_{T1,2}$.

  In order to calculate the spin -dependent transport we need the BC at the injector interface $z=0$ which
  are determined from the general ones (\ref{Eq:BCdf}) and have the form
  \begin{eqnarray} \label{Eq:BCfT121Exact1}
  {\mathcal{D}}_{T1}\nabla f_{T1}+ {\mathcal{D}}_{T2} \nabla f_{T2}
  =\\ \nonumber
  2\kappa_I \{ N_+f_{T1} + {\rm Im} g_{33}[ f_{T2} + P_I (f_T- n_- ) ]  \}
  \\ \label{Eq:BCfT121Exact2}
  {\mathcal{D}}_{T1} \nabla f_{T2} - {\mathcal{D}}_{T2}\nabla f_{T1} =\\ \nonumber
  2\kappa_I \{ N_+ [ P_I( f_T- n_-) + f_{T2} ] -  {\rm Im} g_{33} f_{T1} \} .
  \end{eqnarray}
  These BC are valid for arbitrary values of transparency
  $\kappa_I$. They can be simplified in the spin injection limit $\kappa_I\xi\ll
  1$ where $\xi=\sqrt{D/\Delta}$ is the superconducting coherence
  length. Omitting in the r.h.s. of Eqs.~(\ref{Eq:BCfT121Exact1},\ref{Eq:BCfT121Exact2})
  the terms which are of the second order in $\kappa_I\xi$  we get
   \begin{eqnarray} \label{Eq:BCfT121}
  {\mathcal{D}}_{T1}\nabla f_{T1}+ {\mathcal{D}}_{T2} \nabla f_{T2} &=& - 2\kappa_I {\rm Im} g_{33} P_In_-   \\
  \label{Eq:BCfT122}
  {\mathcal{D}}_{T1} \nabla f_{T2} -  {\mathcal{D}}_{T2}\nabla f_{T1} &=& - 2\kappa_I  N_+ P_I n_-  .
  \end{eqnarray}

  At first glance, both the kinetic Eqs.~(\ref{Eq:KinfT12-1},\ref{Eq:KinfT12-2}) and the BC
  (\ref{Eq:BCfT121Exact1},\ref{Eq:BCfT121Exact2}) have a rather complicated form. However their physical meaning is quite
  clear as we discuss in what follows.
  First,  let us note that the kinetic Eqs.~(\ref{Eq:KinfT12-1},\ref{Eq:KinfT12-2}) can
  be written in a more compact form similar to the
   Landau-Lifshitz-Gilbert equation for a damped gyromagnetic
   precession for the spectral density of spin polarization (\ref{Eq:Magnetization})
   \begin{equation}
   \label{LL}
   \bm{ \nabla\cdot j_s} = g_S{\bm m}_\perp\times ({\bm h} + {\bm  h}_s) + {\bm m}_\perp/\tau_\perp,
   \end{equation}
   where we have introduced  an electronic spin g-factor $g_S=2$ and the transversal spin current density
   \begin{equation}\label{Eq:SpinCurrent}
   \bm{ j_s}=\frac{D({\mathcal{D}}_{T1}+i {\mathcal{D}}_{T2} \sigma_2)}{2}\nabla{\bm
   f}_T.
   \end{equation}
  In Eq.~\eqref{LL} the {\it transverse spin relaxation time} $\tau_\perp$ is given by
  \begin{equation}\label{Eq:SpinRelLLG}
  \tau^{-1}_\perp=  \frac{2h(H_2S_{T1} + H_1S_{T2})}{H_1^2+H_2^2}.
   \end{equation}
   The time $\tau_\perp$ determines the rate of spin coherence
   relaxation in the diffusive superconductor \cite{MakiEPR}. In contrast  to the normal metal where it  coincides with the usual spin relaxation time
 $\tau_\perp=\tau_{sn}$ in the superconducting state $\tau_\perp$ has a
   pronounced energy dependence shown in the right panels of  Fig.~(\ref{Fig:DOm}).
   The spin relaxation terms, $S_{T1,T2}$, not only determine the decay of spin coherence but also
 introduce a shift of the precession frequency which can be expressed as a renormalization of the Zeeman field
  [see the first term on the r.h.s. of Eq. (\ref{LL})]:
  \begin{equation}\label{Eq:ZemanCorrection}
  \bm {h}_s = \bm {h} \frac{H_2S_{T2} - H_1S_{T1}}{H_1^2+H_2^2}\; .
  \end{equation}
 Also this term is energy dependent, as shown  in the  right panels of  Fig.(\ref{Fig:DOm}).

The BC (\ref{Eq:BCfT121Exact1},\ref{Eq:BCfT121Exact2}) can be also
written in a simple form using  the definition of the spin current
(\ref{Eq:SpinCurrent}). We find
 \begin{equation}\label{Eq:BCfT}
 {\bm j_s}|_{z=0}  =  D\kappa_I \left( {\bm m}_\perp|_{z=0} - {\bm m}_N \right),
 \end{equation}
 where ${\bm m}_N = n_-[ N_+ {\bm {P}}_I  + h^{-1}{\rm Im} g_{33}( {\bm{P}_I\times \bm h })]$
 is the spin polarization acquired by non-equilibrium electrons
 which tunnel through the
 spin-filtering ferromagnetic barrier from the voltage-biased normal electrode to the superconductor.
 The above expression for ${\bm m}_N$ is given by the
 Eq.~(\ref{Eq:Magnetization}) with ${\bm f}_T = {\bm P}_I n_-$.
 The BC in spin injection limit
 (\ref{Eq:BCfT121},\ref{Eq:BCfT122}) are obtained by setting ${\bm m}_\perp|_{z=0}=0$
 in the r.h.s of Eq.(\ref{Eq:BCfT}).

Combining
 Eq.~\eqref{Eq:Magnetization} and Eq.~\eqref{Eq:BCfT} one can
 easily verify that there is a finite subgap contribution to the
 spectral spin current originating from the second term in the r.h.s.
 of Eq.~\eqref{Eq:Magnetization}.  It is important to emphasize
 that this subgap spin imbalance appears in linear order in
 $\kappa_I$, and exists exclusively in the presence of a Zeeman
 field and noncollinear spin injection, ${\bm h} \nparallel {\bm P}_I$.

 \section{Hanle effect} \label{Sec:HanleEffect}

\begin{figure}[!htb]
 \centerline{\includegraphics[width=1.0\linewidth]{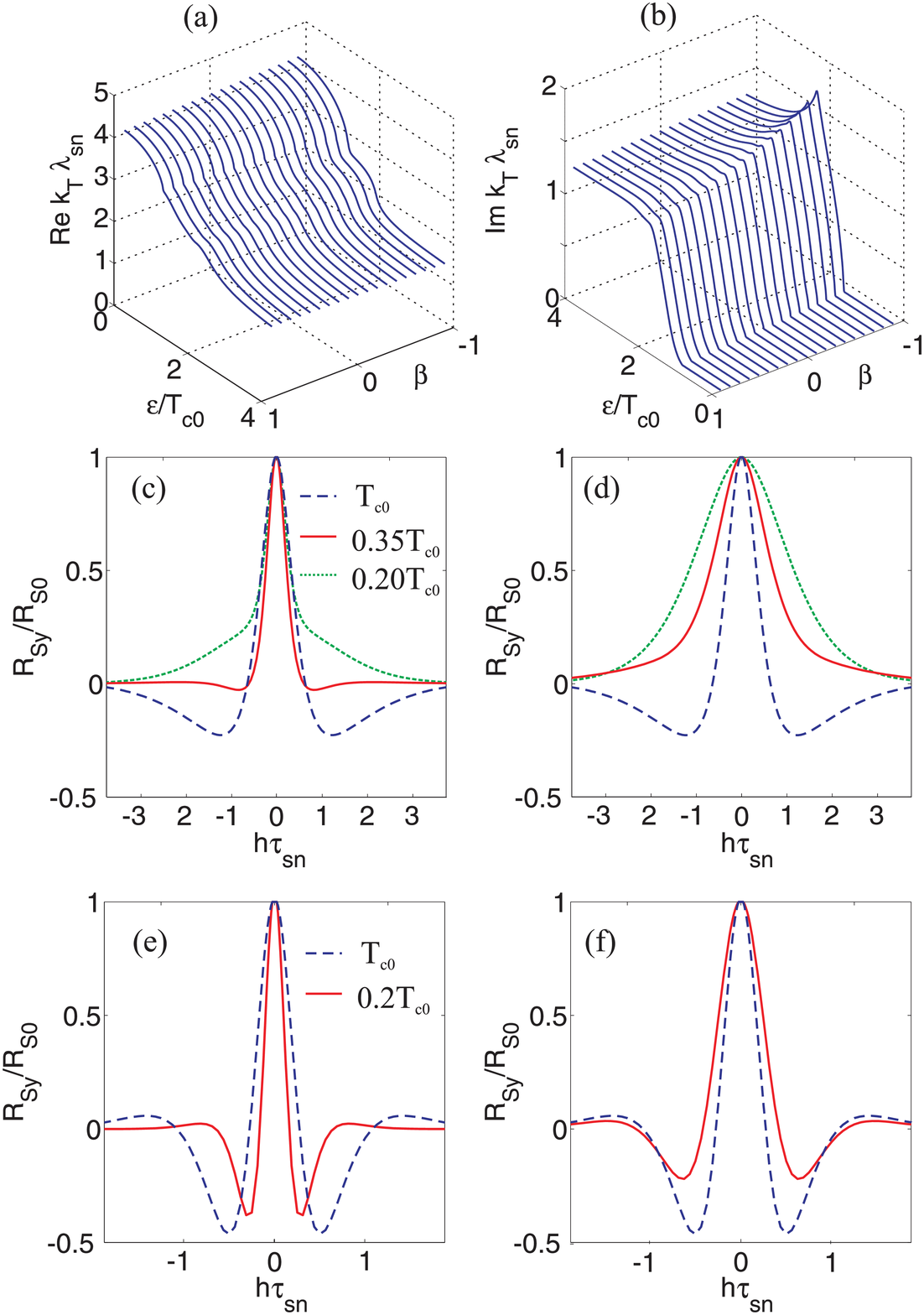}}
 \caption{\label{Fig:Scales} (Color online)
 Energy dependence of the inverse length scales for the (a) decay
 ${\rm Re} k_T$ and (b) oscillations ${\rm Im} k_T$ of the spin imbalance
 for different values of  $\beta$, and $h=1.87/\tau_{sn}$, $T=0.1T_{c0}$.
 The spin-dependent nonlocal magnetoresistance $R_{Sy} (H)$ is shown in
 (c) for  $L_D=2\lambda_{sn}; \; \beta=-0.9 $,  (d) for $L_D=2\lambda_{sn}; \; \beta=0.5$,
 (e) $L_D=5\lambda_{sn}; \; \beta=-0.9$, (f) $L_D=5\lambda_{sn}; \; \beta=0.5$.
 Here $\lambda_{sn} = 1/\sqrt{D\tau_{sn}}$ is the normal state spin relaxation length.  In all panels
 $\tau_{sn} = 5/T_{c0}$. The plots of $R_{Sy} (H)$ are normalized to the absolute maximum values $R_{S0}$ at each temperature. }
 \end{figure}

 Let us now apply the general theory developed in the previous
 section to study the Hanle effect in the model of a non-local spin
 valve discussed in Sec.~(\ref{Sec:Model}). In this effectively 1D system with a spatially homogeneous
distribution of the order parameter Eq.~\eqref{LL}
   can be solved analytically. The components of  ${\bm f}_T$ have the form   $ f_{T1} = -  {\rm Im}( A e^{- k_{T} z } )$   and
   $f_{T2} =   {\rm Re}( A e^{- k_{T}z } )$, where $A$ is an
    integration constant determined by the BC, and
   \begin{equation}\label{Eq:kT}
   k_T= \left[\frac{(S_{T1}-H_1)-i(S_{T2}+H_2)}{D({\mathcal{D}}_{T1}-i{\mathcal{D}}_{T2})}\right]^{1/2}
  \end{equation}
  is the inverse of the characteristic  length with ${\rm Re}k_T>0$. Its real part determines the inverse
  spin relaxation length which is energy dependent. Its imaginary part describes
  the precession  of the spin of quasiparticles with energy $\varepsilon$.
  As can be seen from Figs.~\ref{Fig:Scales}(a),(b), the precession and relaxation
  lengths depend on the nature of the spin scattering mechanism.
  At intermediate temperatures below $T_c$, the main contribution to the spin-dependent ${\bm \mu}$ in Eq.~\eqref{Eq:SpinSignal} comes from energies
  close to the spectral gap $\Delta_g$.
  In the case of dominating spin-orbit scattering ($\beta<0$), one
  clearly sees in Fig.~\ref{Fig:Scales}(a),(b) that while ${\rm Re}\,k_T$
  at $\varepsilon\approx\Delta_g$ is larger than in the normal case ($\varepsilon\gg \Delta_g$),
  the imaginary part of $k_T$  has a peak.
  This results in  a modified Hanle curve for temperatures below $T_c$ [see Figs.~\ref{Fig:Scales}(c),(e)] in which the suppression of the spin
  signal ($R_{Sy}$ corresponding to the detector polarization ${\bm P}_D=P_D {\bm y}$ so that ${\bm P}_D\parallel {\bm P}_I \perp {\bm h}$)
   appears at smaller magnetic fields than in the normal state, while the
   oscillation becomes smaller.
In contrast, if the spin-flip mechanism dominates  ($\beta>0$),
 the imaginary part of $k_T$ is suppressed at
$\varepsilon\approx\Delta_g$ [Figs.~\ref{Fig:Scales}(b)]. This
leads to an increase of the oscillation period of $R_{Sy}(h)$ when
the temperature is decreased [Figs.~\ref{Fig:Scales}(d),(f)]. The
real part ${\rm Re} k_T$ has a larger value than in the normal
state [Fig.~\ref{Fig:Scales}(a)]. Such an increase is mainly
determined by the large renormalization  of the spin relaxation in
the superconducting state associated with spin-flip
scattering\cite{morten04,morten05}.
 In this case the spin
relaxation length has a weaker dependence on the external magnetic
field than in the normal state. This explains the Hanle curves
shown in Figs.~\ref{Fig:Scales}(d),(f),
 where the decay scale of $R_{Sy}(h)$ increases
 towards lower temperatures at $T<T_c$.

 At sufficiently low temperatures, the main contribution to the nonlocal
resistance comes from energies below the gap in
Eqs.~(\ref{Eq:Magnetization},\ref{Eq:BCfT}), and the transport is
dictated by subgap tunneling. This sub-gap process is different
from the previously discussed one \cite{ZaikinCAR}. It exists only
in the presence of a Zeeman field non-collinear with the injector
electrode polarization. In such a system the subgap spin imbalance
in the absence of a charge imbalance appears due to the
gyromagnetic precession of the injected quasiparticle spins driven
by the non-collinear Zeeman field. When the spin-polarized
detector is connected, it converts the spin imbalance into the
charge current according to the Eq.(\ref{Eq:ZeroCurrentYGen}). We
neglect the corrections to the distribution functions which
describe this subgap charge current in the
 superconducting wire.

\begin{figure}[!htb]
 \centerline{\includegraphics[width=1.0\linewidth]{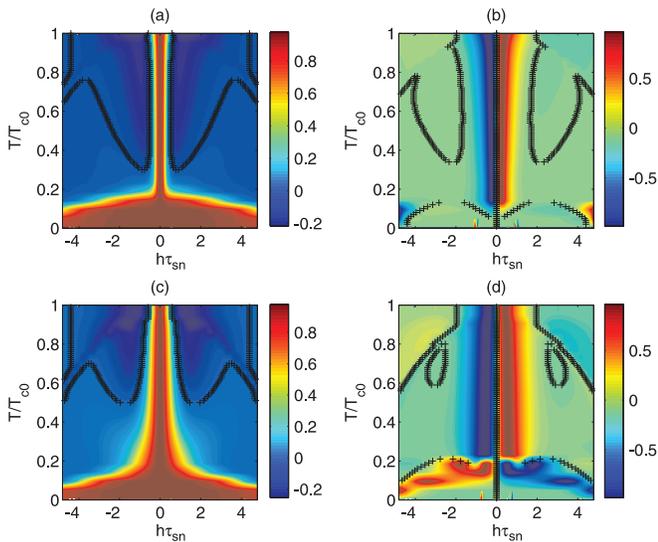}}
 \caption{\label{Fig:Hanle} (Color online)
  Temperature and field dependencies of the nonlocal spin signal. Panels  (a,c) show  $R_{Sy}$
  for  ${\bm P}_D\parallel {\bm y}$ and panels
 (b,d) show $R_{Sx}$ for ${\bm P}_D\parallel {\bm x}$.
 In (a,b) $\beta=-0.9$ and (c,d) $\beta=0.5$.
 The plots are normalized to the absolute maximum values at each temperature.
 The  distance between the injector and the detector is $L_D= 2\lambda_{sn}$,
  $\tau_{sn} = 5/T_{c0}$ and  $\kappa_I \sqrt{D/T_{c0}} =0.001$.  The lines of zeros in $R_{Sx},R_{Sy}$ are shown by black crosses. }
 \end{figure}

The subgap tunneling of quasiparticles between the injector and
detector electrodes leads to a complete elimination of the
coherent spin rotation. Indeed as shown in
Figs.~\ref{Fig:Scales}(a),(b), the inverse length $k_T$ for
$\varepsilon<\Delta_g$ is real. Moreover, it has only a very weak
dependence on $h$. As a result, in the configuration ${\bm
P}_D\parallel {\bm P}_I \perp {\bm h}$ both the precession and
decay of the nonlocal signal disappear at $T \rightarrow 0$, as
shown in Figs.~\ref{Fig:Hanle}(a),(c).

 The threshold temperature of the Hanle effect
suppression is determined by the competition between the
contributions of propagating quasiparticles at the energies above
the gap $\varepsilon>\Delta_g$ and the subgap tunneling process
at $\varepsilon<\Delta_g$. The former has an amplitude of the
order $e^{- L_D/\lambda_{s}} e^{-\Delta_g/T}$ which is
proportional to the density of thermal quasiparticles and decays
over the spin relaxation length in the normal state $\lambda_s\approx \lambda_{sn}$. The
subgap contribution is determined by an almost
temperature-independent factor $e^{- L_D / \xi} $ which decays
however over a  much shorter  length $\xi = \sqrt{
D/\Delta}\ll \lambda_{sn}$ [see the energy dependence of length
scales in Fig.(\ref{Fig:Scales})a].

If instead of the above analyzed configuration one assumes that
the three vectors $({\bm P}_D, {\bm P}_I,{\bm h} )$ are
perpendicular to each other ({\it e.g.} ${\bm P}_D = P_D {\bm x}$,
${\bm P}_I = P_I {\bm y}$, ${\bm h} = h {\bm z}$), the subgap
current is absent in the detector circuit and the corresponding
spin signal $R_{sx}$ has a strong dependence on $h$ even in the
limit $T\rightarrow 0$ (see  Figs. \ref{Fig:Hanle} b,d). Note that
$R_{Sx}(h)=-R_{Sx}(-h)$ and therefore, in contrast to the usual
Hanle effect, this signal can be measured without changing the
magnetization of the electrodes ${\bm P}_D, {\bm P}_I$ but just  by
changing the sign of the external magnetic field.

 \section{Conclusion} \label{Sec:Conclusion}

In conclusion, we have developed a theoretical framework to study spin rotation and relaxation in superconductors in the case of
noncollinear spin fields.
We have analyzed the Hanle effect in a mesoscopic superconductor and demonstrated that the nonlocal magnetoresistance
 deviates  from the one in the normal state.
 Moreover, we show that the Hanle curves depend on the nature of the spin scattering mechanism, either spin-orbit or spin-flip impurities.
 Our findings  provide a way to identify these mechanisms by standard magnetoresistance measurements in nonlocal spin valves,
 and establish the fundamental physics underpinning the spin control and manipulation in superconducting devices.

 \section{Acknowledgments}

The  work of T.T.H  was supported by the Academy of Finland and
the European Research Council (Grant No. 240362-Heattronics)
programs. The work of F.S.B. has been supported by the Spanish
Ministry of Economy and Competitiveness under Project No.
FIS2011-28851-C02-02. F.S.B. acknowledges funding from the
Leverhulme Trust through an International Network Grant (grant
IN-2013-033). P.V. acknowledges the Academy of Finland for
financial support.


\begin{thebibliography}{99}

\bibitem{Spintronics}
I. {\ifmmode \check{Z}\else \v{Z}\fi{}uti\ifmmode \acute{c}\else
\'{c}\fi{}}, J. Fabian, and S. Das Sarma, Rev. Mod. Phys. {\bf
76}, 323 (2004). 

 \bibitem{AwshalomReview}
 D. D. Awschalom, M. E. Flatt\'{e},
Nature Phys. {\bf 3}, 153, (2007)

 \bibitem{GaAsSpinRot}
 J.M. Kikkawa, D.D. Awschalom, Nature, {\bf 397}, 139 (1999). 

 \bibitem{SiliconSpinRot}
 I. Appelbaum, B. Q. Huang, and D. J. Monsma, Nature
 (London) {\bf 447}, 295 (2007). 

\bibitem{GraphenSpinRot}
  N. Tombros, C. Jozsa, M. Popinciuc, H. T. Jonkman, and B. J.
  van Wees, Nature (London) {\bf 448}, 571 (2007). 

\bibitem{Johnson1985} M. Johnson and R.H. Silsbee, Phys. Rev. Lett. {\bf 55}, 1790 (1985).

 \bibitem{Jedema} F.J. Jedema, H.B. Heersche, A.T. Filip, J.J.A. Baselmans,
 B.J. van Wees, Nature {\bf 416}, 713 (2002). 

\bibitem{Jedema1}
 F.J. Jedema, A.T. Filip  and  B.J. van Wees, Nature {\bf 410},
 345 (2001). 

\bibitem{Fukuma2011} Y. Fukuma {\it et al}, Nat. Mat. {\bf 10}, 527
(2011). 

\bibitem{NazarovNonCollinear}
  D. H. Hernando, Yu.V. Nazarov, A. Brataas, and G. E.W.
 Bauer, Phys. Rev. B {\bf 62}, 5700 (2000). 

\bibitem{HanleSuper}
H. Yang, S.-H. Yang, S. Takahashi, S. Maekawa, and S. S. P.
Parkin, Nature Mat. {\bf 9}, 586 (2010). 

\bibitem{Hubler} F. H\"ubler, M.J. Wolf, D. Beckmann, H.v. L\"ohneysen,
 Phys. Rev. Lett. {\bf 109}, 207001 (2012). 

\bibitem{Aprili} C.H.L. Quay, D. Chevallier, C. Bena, M. Aprili,
 Nature Phys. {\bf 9}, 84 (2013). 

\bibitem{ScientificReports}
K. Ohnishi, Y. Ono, T. Nomura, T. Kimura, Sci. Rep. {\bf 4}, 6260
(2014).

\bibitem{EschrigPhysToday}
M. Eschrig, Phys. Today {\bf 64}, 43 (2011).

\bibitem{BergeretNatureComm2014}
N. Banerjee, C. B. Smiet, R. G. J. Smits, A. Ozaeta, F. S.
Bergeret, M. G. Blamire,  J. W. A. Robinson,  Nature Comm. {\bf
5}, 3048 (2014)

\bibitem{morten04} J. Morten, A. Brataas, and W. Belzig, Phys. Rev. B {\bf 70}, 212508
(2004). 

 \bibitem{morten05} J. Morten, A. Brataas, and W. Belzig, Phys. Rev. B {\bf 72}, 014510
(2005). 

\bibitem{Poli} N. Poli, J.P. Morten, M. Urech, A. Brataas, D.B. Haviland, V. Korenivsky,
 Phys. Rev. Lett. {\bf 100}, 136601 (2008). 

 \bibitem{SpinChargeSeparation}
 H. L. Zhao and S. Hershfield, Phys. Rev. B {\bf 52}, 3632 (1995).

 \bibitem{SpinInjectionNb}
 T. Wakamura, N. Hasegawa, K. Ohnishi, Y. Niimi, and YoshiChika Otani,
 Phys. Rev. Lett. {\bf 112}, 036602 (2014).

\bibitem{HekkinkCAR}
G. Falci, D. Feinberg and F. W. J. Hekking, Europhys. Lett., {\bf
54}, 255 (2001). 

  \bibitem{BeckmanCAR}
D. Beckmann, H. B. Weber, H. v. Lohneysen, Phys. Rev. Lett. {\bf
93}, 197003 (2004). 

  \bibitem{ZaikinCAR}
M.S. Kalenkov, A.D. Zaikin, Phys. Rev. B {\bf 82}, 024522 (2010). 

 \bibitem{SpinSignalDivergence}
  S. Takahashi and S. Maekawa, Phys. Rev. B {\bf 67}, 052409
(2003). 

  \bibitem{Bergeret2001}
 F. S. Bergeret, A. F. Volkov, and K. B. Efetov,  Rev. Mod. Phys. {\bf 77}, 1321
(2005). 

  \bibitem{AbrikosovGorkov}
 A. A. Abrikosov, L. P. Gor'kov, Zh. Eksp. Teor. Fiz. {\bf 39}, 1781 (1960)
  [Sov. Phys. JETP {\bf 12}, 1243 (1961)]. 

\bibitem{SaintJames}
D. Saint-James, D. Sarma, and E. J. Thomas, {\it Type II
Superconductivity}, Pergamon, New York (1969). 

\bibitem{SchmidSchon}
A. Schmid, G. Sch\"on, J. Low. Temp. Phys. , {\bf 20}, 207 (1975)

\bibitem{MakiEPR}
K. Maki, Phys. Rev. B {\bf 8}, 191 (1973). 

 \bibitem{bergeret12} F.S. Bergeret, A. Verso, and A.F. Volkov, Phys. Rev. B {\bf 86}, 214516
(2012). 

\bibitem{KL} M.Y. Kuprianov and V.F. Lukichev, Sov. Phys. JETP {\bf 67}, 1163
(1986). 

\end{thebibliography}
\end{document}